\documentclass[journal]{IEEEtran}
\usepackage{amsmath,amsfonts}
\usepackage{array}
\usepackage[caption=false,font=normalsize,labelfont=sf,textfont=sf]{subfig}
\usepackage{textcomp}
\usepackage{stfloats}
\usepackage{url}
\usepackage{verbatim}
\usepackage{graphicx}
\usepackage{cite}
\usepackage{xcolor}
\usepackage{booktabs}
\usepackage[most]{tcolorbox}
\usepackage{multirow}
\usepackage{enumitem}
\usepackage{siunitx}
\usepackage{hyperref}
\definecolor{LightBlue}{RGB}{102,153,255}
\definecolor{DeepYellow}{RGB}{204,153,0}
\definecolor{DeepGreen}{RGB}{0,110,60}

\newcommand{\tok}[1]{\texttt{<#1>}}
\newcommand{\tsep}{\penalty0\hspace{0pt}}
\newcommand{\itsep}{\texttt{, }\penalty0\hspace{0pt}}
\newcommand{\up}[1]{\,{\scriptsize\textcolor{red}{$\uparrow$\,#1}}}
\newcommand{\down}[1]{\,{\scriptsize\textcolor{blue}{$\downarrow$\,#1}}}
\newcommand{\same}[1]{\,{\scriptsize\textcolor{gray}{$\rightarrow$\,#1}}}

\begin{document}

\title{Position-Aware Drafting for Inference Acceleration in LLM-Based Generative List-Wise Recommendation}

\author{
Jiaju~Chen$^{*,\dagger}$, Chongming~Gao$^{*,\S}$, Chenxiao~Fan$^{*}$, Haoyan~Liu$^{*}$, Qingpeng~Cai$^{\ddagger}$, Peng~Jiang$^{\ddagger}$, and Xiangnan~He$^{*,\S}$\\
$^{*}$University of Science and Technology of China, Hefei, China\\
$^{\dagger}$Zhongguancun Academy, Beijing, China\\
$^{\ddagger}$Independent Researcher\\
$^{\S}$Corresponding author\\
Email: cjj01@mail.ustc.edu.cn, chongming.gao@gmail.com, simonfan@mail.com, liuhaoyan@ustc.edu.cn,\\
cqpcurry@gmail.com, jp2006@139.com, xiangnanhe@gmail.com
}

\markboth{IEEE Transactions on Knowledge and Data Engineering,~Vol.~XX, No.~XX, XXXX~2026}{Chen \MakeLowercase{\textit{et al.}}: Position-Aware Drafting for Inference Acceleration in LLM-Based Generative Recommendation}

\maketitle

\begin{abstract}
Large language model (LLM)–based generative list-wise recommendation has advanced rapidly, but decoding remains sequential and thus latency-prone. To accelerate inference without changing the target distribution, \emph{speculative decoding} (SD) uses a small \emph{draft model} to propose several next tokens at once and a target LLM to verify and accept the longest prefix, skipping multiple steps per round. In generative recommendation, however, each item is represented by multiple semantic-ID tokens (often with separators), and current drafts typically treat these tokens uniformly. This overlooks two practical facts: (i) a token’s semantics depend on its \emph{within-item slot}, and (ii) uncertainty tends to increase with \emph{speculation depth}. Without modeling these effects, SD’s speedups can be limited.
We introduce PAD-Rec (\textbf{P}osition-\textbf{A}ware \textbf{D}rafting for generative \textbf{Rec}ommendation), a lightweight module that augments the draft model with two complementary signals. \textit{Item position embeddings} explicitly encode the within-item slot of each token, strengthening structural awareness. \textit{Step position embeddings} encode the draft step, allowing the model to adapt to depth-dependent uncertainty and improve proposal quality. To harmonize these signals with base features, we add simple gates—a learnable coefficient for item slots and a context-driven gate for draft steps. The module is trainable, easy to integrate with standard draft models, and adds negligible inference overhead. Extensive experiments on four real-world datasets show up to \textbf{3.1$\times$} wall-clock speedup and about \textbf{5\%} average wall-clock speedup gain over strong SD baselines, while largely preserving recommendation quality. Our code is available at \url{https://github.com/Jiaju-Chen/PAD-Rec}.

\end{abstract}


\section{Introduction}

With the rapid rise of large language models (LLMs), their application in generative list-wise recommendation systems has gained significant momentum, particularly in areas like end-to-end recommendation \cite{chen2025onesearch,deng2025onerec,chen2025dlcrec} and conversational recommendation \cite{zou2024knowledge}.

These systems generate personalized item lists by constructing user profiles and iteratively invoking LLMs in an auto-regressive manner. 
However, the auto-regressive nature of LLMs results in relatively slow generation speeds, leading to high latency and hindering their application in real-time recommendation scenarios~\cite{wang2025decoding}. 
Therefore, accelerating inference in LLM-based generative recommendation systems has become an urgent issue to address \cite{lin2025efficient,xi2025efficiency}. 

\begin{figure}[!t]
\setlength{\abovecaptionskip}{0cm}
\setlength{\belowcaptionskip}{-0.3cm}
\centering
\includegraphics[width=1\linewidth]{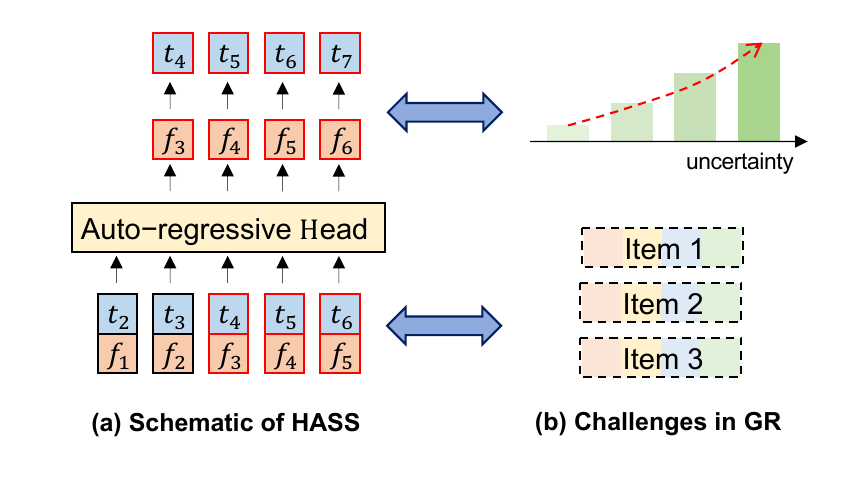}
\caption{(a) Schematic of a HASS-style draft when predicting the 4th to 7th tokens. Blue blocks ($t$) denote tokens and orange blocks ($f$) denote intermediate features; subscripts index sequence positions. Red outlines indicate draft predictions. (b) Challenges in GR: item-structured tokenization and uncertainty that grows with generation depth.}
\label{fig:hass}
\end{figure}

A prominent line of work for {lossless} acceleration of autoregressive LLMs is \emph{speculative decoding} (SD), which follows a \emph{draft--then--verify} scheme \cite{leviathan2023fast,xia2022speculative,xia2024unlocking,chen2023accelerating}. A small, compatible \emph{draft} model proposes several future tokens at once (as a candidate block or tree), and the \emph{target} LLM verifies these candidates in a single batched call. Verification accepts the longest prefix that matches the target distribution; decoding then resumes from that prefix. By committing multiple tokens per target call, SD reduces the number of expensive target invocations while preserving the exact decoding distribution \cite{miao2024specinfer,zhou2023distillspec}.
Building on this paradigm, later studies focus on raising the {acceptance rate}, which directly improves end-to-end speed. EAGLE \cite{li2024eagle} enriches the draft input with token embeddings and \emph{features} (hidden states before the target head), leading to more faithful proposals and higher acceptance; however, its training is \emph{single-step}, whereas inference drafts {multiple} steps, creating a train–decode mismatch. HASS~\cite{zhang2025learning}, as illustrated in Fig.~\ref{fig:hass}\,(a), addresses this gap by training the draft to \emph{roll forward multiple steps}, so the training context mirrors the multi-step drafting used at inference. The common goal is clear: keep the draft fast and well-aligned with the target LLM, achieving larger speedups with SD.

Yet, even with stronger inputs and multi-step training, directly applying SD to \emph{generative list-wise recommendation} exposes two task-specific gaps that prior SD works---including HASS---do not explicitly resolve.
As highlighted in Fig.~\ref{fig:hass}\,(b), recommendation outputs are {item-structured}: each item is emitted as a fixed tuple of $K$ semantic-ID tokens (typically $K{=}4$)~\cite{qu2025tokenrec,zheng2024adapting}, corresponding to different codebook levels in RQ-VAE-style representations~\cite{lee2022autoregressive}.
We refer to these fixed positions as {within-item slots}, indexed by $1{:}K$ in the semantic-ID tuple.
Meanwhile, the {autoregressive} nature of decoding---together with SD’s “stop at first mismatch” verification---tends to make later draft steps harder to be accepted.
Concretely, two challenges arise for the {draft model}:

\textbf{(i) Slot-conditioned semantics are underused.}
Because each item occupies fixed slots, a token’s semantics depend on its {within-item slot}; however, existing draft models treat tokens uniformly and do not encode slot identity, weakening intra-item modeling and alignment with the target head.

\textbf{(ii) Depth-driven uncertainty is unmanaged.}
Independently of the fixed layout, uncertainty often {increases with the draft depth}: later tokens depend on longer histories (error accumulation) and are more likely to trigger early rejection under SD’s verification. Standard training remains largely step-agnostic, so the draft model does not adapt its proposal strategy or confidence by depth, leading to miscalibration and lower acceptance at deeper positions.

To close these gaps, we introduce PAD-Rec—a position-aware acceleration module that plugs into standard draft models and directly targets the two gaps above. The draft model is equipped with two complementary positional signals: \emph{item position embeddings} (IPE) that encode a token’s slot within the item to strengthen item-structure awareness, and \emph{step position embeddings} (SPE) that index the current draft step so the draft can learn depth-conditioned behavior from data and adapt to step-wise uncertainty. To balance these signals with base features, we use simple gates: a learnable coefficient for IPE and a small, context-driven gate for SPE. The module is lightweight, trained jointly with the draft, easy to integrate, and adds negligible inference overhead—yielding higher verification acceptance and larger end-to-end speedups.

Extensive experiments on four real-world datasets demonstrate that our method achieves up to \textbf{3.1$\times$} speedup and an average \textbf{5\%} wall-clock speedup gain over HASS in LLM-based generative list-wise recommendation, while largely preserving recommendation performance.
Beyond the main comparison, we conduct further analyses to examine the contribution of individual components and to study the effects of key hyper-parameters and model scaling.

The contributions of this paper are as follows: 
\begin{itemize}[leftmargin=*]
\item To our knowledge, PAD-Rec is the first to explicitly exploit within-item token positions and draft-step positions for speculative decoding in structured list-wise generative recommendation.
\item We propose a method for accelerating generative recommendation by incorporating item-position and step-position embeddings into the draft model.
\item We demonstrate the effectiveness of our method through extensive experiments, achieving up to \textbf{3.1$\times$} acceleration for LLM-based generative recommendation tasks while largely preserving recommendation performance.
\end{itemize}

\section{Related Work}

\subsection{Speculative Decoding}
Speculative decoding accelerates LLMs via a \emph{draft--then--verify} loop: a small draft model proposes several future tokens, and the target LLM verifies them in one batched call, committing the longest accepted prefix~\cite{leviathan2023fast,chen2023accelerating,xia2024unlocking}. 
Early SD under \emph{greedy} decoding drafts a short block per round, preserving the target distribution but limiting acceptance and speedup~\cite{stern2018blockwise}. Later works adopt nucleus sampling~\cite{leviathan2023fast} and generalize to \emph{tree-based} proposals that branch and verify in parallel, improving acceptance length at the cost of extra memory when trees widen or deepen~\cite{li2024eagle2,miao2024specinfer,chen2024sequoia,svirschevski2024specexec}.
By drafting strategies, SD methods can be grouped into three broad categories. 
(i) \emph{Prompt-lookup} SD retrieves candidate continuations from the input prompt itself via suffix matching and submits them for batched verification, avoiding a separately trained draft model~\cite{fu2024break,kou2024cllms}. 
(ii) \emph{Retrieval-based} SD augments the prompt with contexts retrieved from external stores or prior outputs, leveraging cached knowledge to guide drafting~\cite{he2024rest,zhao2024ouroboros}. 
(iii) \emph{Lightweight-drafter} SD attaches efficient predictors to generate candidates, including token-head~\cite{cai2024medusa}, RNN-based models~\cite{ankner2024hydra}, or smaller models~\cite{lin2025efficient} for parallel or regressive proposals.
Recently, the third group has given rise to \textit{feature-based} SD. EAGLE~\cite{li2024eagle} is a representative approach that autoregresses over hidden-state sequences to fuse candidate token embeddings. This line has inspired follow-ups such as HASS~\cite{zhang2025learning}, GRIFFIN~\cite{hu2025griffin}, FSPAD~\cite{gui2024boosting}, and CORAL~\cite{weng2025coral}, which refine draft-model training and proposal construction.
Other studies examine broader efficiency questions, including the trade-off between draft-model size, draft length, and latency~\cite{liu2024speculative}, context-dependent uncertainty in SD~\cite{liu2025heterospec}, and systems optimizations such as KV-cache reduction~\cite{zhang2023h2o}. However, these methods are mainly designed for generic language generation and do not explicitly model the structured semantic-ID outputs in generative recommendation. This motivates studying structure-aware drafting for list-wise recommendation generation.

\subsection{Generative Recommendation}
Motivated by the rise of generative modeling, recommendation has been reframed from \emph{scoring} to \emph{generation}~\cite{bao2025bi,shi2024large,zhang2025reinforced}. A widely used blueprint comprises two stages: (i) a \emph{tokenization} stage that maps items to discrete symbols, and (ii) a \emph{sequence generation} stage that predicts future items conditioned on user history. As a representative, TIGER~\cite{rajput2023recommender} encodes each item as a short tuple of semantic-ID tokens, typically learned via RQ-VAE~\cite{lee2022autoregressive}, and uses an encoder--decoder to generate recommendation lists. On the tokenization side, recent work trains more expressive and well-aligned codebooks~\cite{hou2025actionpiece,wang2024learnable,qu2025tokenrec,hua2023index}; for instance, LETTER integrates both collaborative and contextual priors to guide semantic IDs~\cite{wang2024learnable}. On the generation side, methods enhance how user profiles and item signals are fused for list-wise decoding~\cite{geng2022recommendation,zheng2024adapting,wang2024eager,zhai2024actions}, exemplified by EAGER's two-stream fusion of semantic and collaborative cues~\cite{wang2024eager} and LC-Rec's instruction-driven generation over semantic IDs~\cite{zheng2024adapting}. Recent studies further apply planning, reasoning, reinforcement learning, and alignment techniques to generative recommendation~\cite{shi2024large,zhang2025reinforced,gao2025sprec,gao2025process}. Personalization is also emerging as a clear trend in generative recommendation and related LLM-based user modeling settings~\cite{zhao2025nextquill,zhao2026don,wang2025think,Gao_survey_2026}. Some systems optimize both stages jointly~\cite{si2024generative,tan2024idgenrec,deng2025onerec}, such as SEATER, which learns a hierarchical tree-structured tokenizer together with an autoregressive encoder--decoder, and IDGenRec, which learns concise textual IDs within the LLM vocabulary for generation. Despite these advances, most methods primarily improve modeling, alignment, or personalization, while the latency of autoregressively generating semantic-ID lists remains underexplored.

\subsection{Inference Acceleration for LLM-based Generative Recommendation}
LLM-based generative recommenders~\cite{zheng2024adapting,qu2025tokenrec} are increasingly competitive with strong traditional systems~\cite{cheng2016wide,he2020lightgcn}, but real-time deployment remains constrained by the latency of autoregressive decoding over multi-token item representations. Beyond compression-based acceleration such as distillation, pruning, and quantization~\cite{hassibi1992second,zheng2025boosting,xiao2023smoothquant}, recent work has begun to improve efficiency more directly in recommendation scenarios. DARE~\cite{xi2025efficiency} accelerates LLM-generated user or item knowledge for downstream ranking rather than directly generating recommendation lists. AtSpeed~\cite{lin2025efficient} studies SD for top-$K$ item generation, where beam search produces multiple candidate semantic-ID sequences and strict $N{\to}K$ verification is required. SpecGR~\cite{ding2026inductive} formulates speculation as retrieval-based candidate proposal and verification for inductive recommendation. NEZHA~\cite{wang2026nezha} uses self-drafting and efficient verification for industrial candidate generation. EARN~\cite{yang2025earn} reduces inference overhead from a systems/cache perspective under item-identifier generation settings.
These studies broaden the design space of efficient generative recommendation, but they mainly target ranking, top-$K$/candidate-generation-oriented settings, retrieval speculation, or cache management rather than full list-wise decoding. In such settings, the generated ``list'' is typically a set of alternative candidate items produced from the same user-history prefix, whereas our setting generates an ordered multi-item response in which previously generated items become part of the decoding context. This full-list setting introduces additional structure: tokens have fixed within-item roles, and later draft steps must account for both deeper speculation and previously generated list tokens. This motivates our focus on structure-aware lossless speculative decoding for full recommendation-list generation.

\section{Preliminary}
\subsection{LLM-based Generative Recommendation}
\label{sec2.1}

We focus on the setup for LLM-based generative recommendation, where a user's interaction history is split into a \emph{history} and a \emph{future}. The history, denoted as $\boldsymbol{a}_{1:n}=(a_1,\dots,a_n)$, consists of items $a_i$ that the user has interacted with, and the future, denoted as $\boldsymbol{b}_{1:m}=(b_1,\dots,b_m)$, is the target list of items to generate.

In addition to item history, the LLM input includes an \emph{instruction} describing the task~\cite{bao2025bi,zheng2024adapting} (e.g., “Given the user's past interactions, predict the list of items the user is most likely to interact with in the future.”).

Following prior work~\cite{rajput2023recommender,zheng2024adapting,wang2024learnable}, we encode items with an RQ-VAE~\cite{lee2022autoregressive} so they can be handled by a token-based generator. Concretely, each \emph{item} is mapped to a short tuple of \emph{semantic ID tokens}. With $K$ codebooks (we use $K{=}4$), item $a_i$ is represented as $(a_i^{(1)},\ldots,a_i^{(K)})$ and item $b_j$ as $(b_j^{(1)},\ldots,b_j^{(K)})$, where the superscript indexes the codebook level. To separate adjacent items, we insert special \emph{separator tokens} (commas/spaces).

\paragraph{Flattened token stream.}
Let $X=(x_1,\dots,x_T)$ be the \emph{textual} token stream that concatenates (i) the instruction and context tokens, (ii) the history items encoded as their $K$ semantic-ID tokens with separators, and (iii) the future items in the same format. Let $t_0$ be the index of the first token of the future (response) segment. Generation proceeds autoregressively over $X$:
\begin{equation}
\label{eq:ar_flat}
p_\theta\big(X_{t_0:T}\mid X_{1:t_0-1}\big)
= \prod_{t=t_0}^{T} p_\theta\big(x_t \mid X_{1:t-1}\big),
\end{equation}
where each $x_t$ ($t\ge t_0$) is a semantic-ID token or separator token, and $X_{1:t_0-1}$ denotes the instruction and history prefix.

\paragraph{Latency implication.}
Autoregressive decoding requires one forward pass per token and is strictly sequential. For a list of length $m$ with $K$ semantic-ID tokens per item and $K'$ separator tokens per item on average, the response segment contains approximately $m\,(K{+}K')$ tokens (in addition to the instruction/history prefix), leading to $m\,(K{+}K')$ decoding steps. This limits parallelism and increases latency as models scale, motivating inference acceleration tailored to generative recommendation.

\subsection{Speculative Decoding and Its Challenges in Generative Recommendation}

To reduce autoregressive latency, \emph{speculative decoding} (SD) \cite{leviathan2023fast} employs a small \emph{draft} model to propose multiple future tokens in parallel as a \emph{candidate block}—or a \emph{candidate tree} in tree-based variants—after which the target LLM scores these proposals and accepts a target-verified prefix, allowing decoding to advance by multiple tokens in one round. Subsequent work refines this pipeline \cite{li2024eagle,li2024eagle2,zhang2025learning,hu2025griffin}. \textbf{EAGLE} \cite{li2024eagle} strengthens the draft by augmenting it with \emph{candidate token embeddings} (plus contextual features) so proposals better match the target head and thus improve acceptance, but its training is \emph{single-step} while inference drafts \emph{multiple} steps—creating a train–decode mismatch. \textbf{HASS} \cite{zhang2025learning} addresses this by training the draft model to roll forward multiple steps so the training context mirrors multi-step drafting at test time. For clarity going forward, we denote by \(B\) the \emph{speculation depth}, i.e., the number of draft steps proposed per round; we unroll up to \(B\) steps during training and expand candidates up to depth \(B\) at inference, and—unless otherwise noted—use the same depth for both (\(B_{\text{train}}{=}B_{\text{test}}{=}B\)).

While these advances raise acceptance in general language tasks, directly applying SD to {LLM-based generative recommendation} exposes two task-specific gaps. First, \emph{slot-positional structure is underused}: each item is realized as a fixed tuple of semantic-ID tokens (typically four IDs plus separators), and a token’s role depends on its {within-item slot}, yet standard draft models treat tokens uniformly without encoding slot identity, weakening intra-item modeling and alignment. Second, \emph{depth-driven uncertainty is unmanaged}: later draft steps—both across items and within an item’s token tuple—are typically harder, but the draft model’s behavior is largely step-agnostic, missing the opportunity to adapt proposal confidence as depth grows.

These observations motivate augmenting the draft model with lightweight position signals that encode the within-item slot and the draft step. We introduce such a position-aware drafting module in the next section.

\section{Method}

\begin{figure}[!t]
\setlength{\abovecaptionskip}{2pt}
\setlength{\belowcaptionskip}{0pt}
\centering
\includegraphics[width=\linewidth]{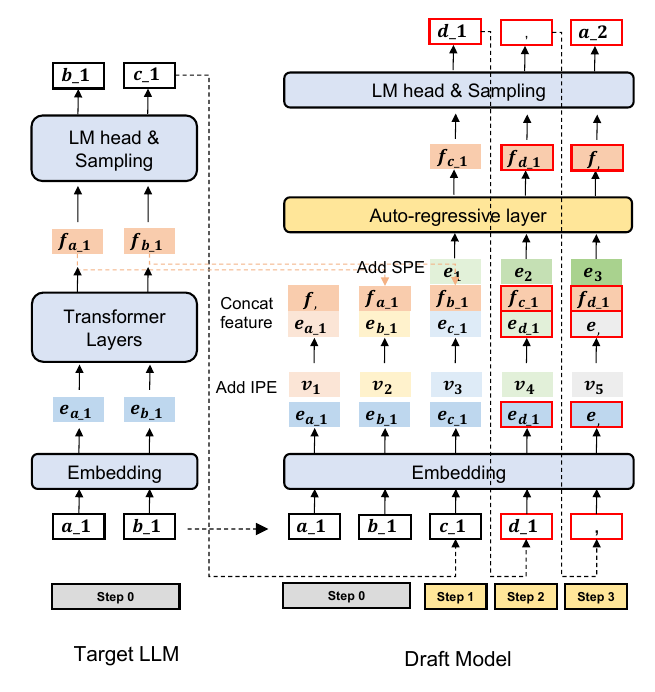}
\caption{Training framework of PAD-Rec. \textcolor{blue}{Blue} blocks denote token embeddings, and \textcolor{orange}{orange} blocks denote intermediate features. Tokens already verified by the target LLM are outlined in \textcolor{black}{black}, while draft predictions are outlined in \textcolor{red}{red}. The draft input is augmented by adding IPE $\mathbf{v}_t$, concatenating features, and adding SPE $\mathbf{s}_j$ before the auto-regressive draft layer. Different IPE colors indicate different within-item positions, while darker SPE colors indicate larger draft-step indices \(j\).}
\label{fig:framework}
\end{figure}

In this section, we address the two challenges identified for SD in LLM-based generative recommendation—underused item-structural cues and unmanaged depth-driven uncertainty. We propose a \emph{position-aware drafting module} that injects two complementary signals into the draft model: (i) \textit{item position embeddings} to encode the within-item slot, and (ii) \textit{step position embeddings} to encode the generation depth. Fig.~\ref{fig:framework} illustrates the integration of the two embeddings in PAD-Rec. To balance these signals with the base representation, we further introduce \textit{gating mechanisms}: a learnable gate for item-position cues and an uncertainty-aware gate for step-position cues. We first detail each component and its integration with SD, and then outline the overall training/inference pipeline.

\subsection{Item Position Embeddings}

\begin{figure*}[t]
\setlength{\abovecaptionskip}{2pt}
\setlength{\belowcaptionskip}{-2pt}
\centering
\includegraphics[width=\linewidth]{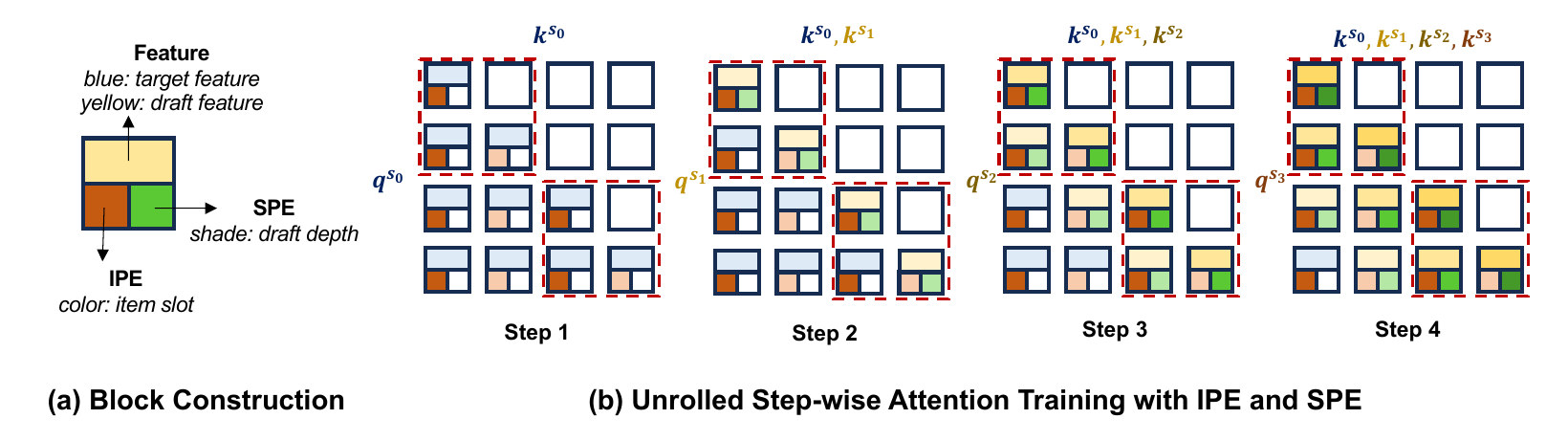}
\caption{\textbf{Position-aware unrolled training with IPE and SPE.}
\textbf{(a)} Each block combines the base feature with an IPE marker for the within-item slot and an SPE marker for the draft step.
\textbf{(b)} During unrolled training, target features are progressively replaced by draft features from earlier depths under a causal mask, while red dashed boxes group tokens belonging to the same item.}

\label{fig:attn}
\end{figure*}

As outlined in Section~\ref{sec2.1}, each item is emitted as a fixed tuple of $K$ semantic-ID tokens, typically $K{=}4$, corresponding to different codebook levels in RQ-VAE~\cite{lee2022autoregressive}.
We index the \emph{within-item slots} as $1{:}K$, while separator tokens and instruction/context tokens are treated as special markers. A typical instruction–response pair looks like:

\begin{tcolorbox}[
  colback=white,
  colframe=black,
  arc=2mm,
  boxrule=0.5pt,
  left=6pt,right=6pt,top=6pt,bottom=6pt,
  breakable,
  before upper=\ttfamily\raggedright
]
\textbf{Instruction:}\\
After interacting with items
\tok{a\_13}\tsep\tok{b\_202}\tsep\tok{c\_9}\tsep\tok{d\_188}\itsep
\ldots\itsep
\tok{a\_171}\tsep\tok{b\_50}\tsep\tok{c\_228}\tsep\tok{d\_186},
what are the next 10 items that could be recommended for the user?

\medskip
\textbf{Response:}\\
\tok{a\_171}\tsep\tok{b\_154}\tsep\tok{c\_190}\tsep\tok{d\_26}\itsep
\tok{a\_13}\tsep\tok{b\_216}\tsep\tok{c\_67}\tsep\tok{d\_214}\itsep \ldots \itsep
\tok{a\_23}\tsep\tok{b\_143}\tsep\tok{c\_82}\tsep\tok{d\_240}
\end{tcolorbox}

To make the slot identity explicit to the draft model, we introduce \emph{item position embeddings} (IPE).
Each token in the flattened decoding stream is associated with a slot or marker label $\ell_t \in \{1,\dots,K,\texttt{sep},\texttt{ctx}\}$, determined by a simple rule based on the known item structure, where \texttt{sep} uniformly denotes both commas and spaces and \texttt{ctx} denotes instruction/context.
The corresponding IPE is obtained by a lightweight embedding lookup:
\begin{equation}
\mathbf{v}_{t} = \mathrm{Emb}^{\text{item}}(\ell_t),
\end{equation}
where $\mathrm{Emb}^{\text{item}}(\cdot)$ is a small embedding table shared across all positions.
By injecting IPE into the draft model, we provide explicit slot-aware cues that strengthen intra-item structural modeling, while adding negligible computational overhead.

\subsection{Step Position Embeddings}

Predictions made at later draft steps are typically less certain due to longer dependency chains and error accumulation.
To enable depth-aware adaptation, we introduce \emph{step position embeddings} (SPE), which explicitly encode the current draft step.

Here, $j$ denotes the draft depth associated with the current token position during speculative decoding. At draft step $j$, the draft model retrieves the corresponding SPE via an embedding lookup:
\begin{equation}
\mathbf{s}_{j} = \mathrm{Emb}^{\text{step}}(j),
\end{equation}
where $\mathrm{Emb}^{\text{step}}(\cdot)$ is a small embedding table indexed by the draft step. In our implementation, draft depth indices start from $j{=}1$, and the SPE table is indexed accordingly.
SPE is applied consistently during both training and inference, providing explicit conditioning on draft depth without introducing additional computation or model complexity.

\subsection{Gating Mechanisms}
\label{sec:gate}

We fuse the two position signals---\emph{item position embeddings} and \emph{step position embeddings}---into the draft model with lightweight gates, so that positional cues strengthen structure/depth awareness without overwhelming the base representations.
Let $\mathbf{e}_{t}\in\mathbb{R}^{d}$ be the token embedding at position $t$, $\mathbf{f}_{t-1}\in\mathbb{R}^{d}$ the draft feature from the previous step, $\mathbf{v}_{t}\in\mathbb{R}^{d}$ the IPE of token $t$, and $\mathbf{s}_{j}\in\mathbb{R}^{d}$ the step position embedding corresponding to the \emph{current draft depth} at position $t$.

\paragraph{Stage-1: slot-aware fusion}
We first inject IPE into the token embedding and concatenate it with the previous-step draft feature:
\begin{equation}
\mathbf{f}'_{t-1}
=
\mathrm{concat}\Big(\mathbf{e}_{t} + g_{\text{item}}\mathbf{v}_{t},\ \mathbf{f}_{t-1}\Big),
\label{eq:gating-stage1}
\end{equation}
where $g_{\text{item}}\in[0,1]$ is a \emph{learnable scalar} that controls the overall strength of slot cues. The concatenation follows the feature-level drafting paradigm of EAGLE~\cite{li2024eagle}, 
which explicitly conditions the current prediction on both token-level embeddings and previously drafted features,
thereby reducing feature uncertainty during speculative decoding.

\paragraph{Stage-2: depth-aware fusion}
We then project the concatenated feature back to the draft-feature space and add SPE with a \emph{context-driven} gate:
\begin{equation}
\mathbf{z}_{t-1}
=
\mathrm{FC}_{\text{cat}}\left(\mathbf{f}'_{t-1}\right),
\quad
\mathbf{z}_{t-1}\in\mathbb{R}^{d},
\label{eq:cat-project}
\end{equation}
\begin{equation}
\mathbf{f}_{t}
=
\mathbf{z}_{t-1} + g_{\text{step}}(t)\mathbf{s}_{j}.
\label{eq:gating-stage2}
\end{equation}
Here, $\mathrm{FC}_{\text{cat}}(\cdot)$ maps the concatenated $2d$-dimensional feature to the $d$-dimensional draft-feature space, so $\mathbf{z}_{t-1}$ and $\mathbf{s}_{j}$ have matched dimensions.

\paragraph{Context-driven step gate.}
Unlike $g_{\text{item}}$, which is global and static, the step gate adapts to the current draft context:
\begin{equation}
g_{\text{step}}(t)=\sigma\!\left(\mathbf{w}^{\top}\mathbf{z}_{t-1}\right),
\label{eq:gating-stepgate}
\end{equation}
where $\sigma(\cdot)$ is the sigmoid function and $\mathbf{w}$ is a learned vector.
This design allows the model to increase the influence of depth cues when drafting becomes more uncertain, while down-weighting SPE when the draft features are confident.

\subsection{Training Pipeline}

PAD-Rec adopts a HASS-style distillation regime with multi-step rollout.
We train the draft model \emph{only on the response part} (semantic-ID tokens and separators) while keeping the target LLM frozen.
The trainable parameters include the draft layer, the IPE/SPE embedding tables, and the gating parameters.

Let $\mathbf{x}_{1:T}$ be the token stream and $t_0$ the index of the first response token.
We define the \emph{speculation depth} $B$ as the maximum number of future tokens that the draft model is allowed to propose in a single speculative round.
For each draft depth $j\in\{1,\dots,B\}$, we distill the draft distribution to match the target distribution on response positions $t\in[t_0,\,T]$:
\begin{equation}
\label{eq:padrec-loss}
\begin{aligned}
\mathcal{L}_{\text{PAD-Rec}}
&= \sum_{j=1}^{B} \Biggl\{ \sum_{t=t_0}^{T}
\mathrm{CE}\Bigl(
P^{(l)}(x_{t}\mid x_{< t}), \\
&\qquad\qquad P^{(s)}_{\theta,\,j}(x_{t}\mid x_{< t})
\Bigr)
+ \text{Aux-loss} \Biggr\}.
\end{aligned}
\end{equation}

$\mathrm{Aux\!-\!loss}$ is the Top-$K$ distillation loss of HASS, which we adopt unchanged.
$P^{(l)}$ denotes the target LLM distribution, and $P^{(s)}_{\theta,\,j}$ denotes the draft distribution conditioned on the \emph{draft depth} $j$ with PAD-Rec’s position-aware inputs:
\begin{equation}
\label{eq:padrec-ps}
P^{(s)}_{\theta,\,j}\left(x_{t}\mid x_{< t}\right)
=\;
\mathrm{Head}\Big(
\mathcal{M}_{\theta}\big(
\mathrm{Fuse}(\mathbf{f}_{< t},\,\mathbf{v}_{< t},\,\mathbf{s}_{j})
\big)
\Big),
\end{equation}
where $\mathcal{M}_{\theta}$ is the trainable draft backbone (a one-layer Transformer block), and $\mathrm{Head}$ is the frozen LM head copied from the target LLM.
$\mathrm{Fuse}(\cdot)$ implements the gated position-aware fusion defined in Eqs.~(\ref{eq:gating-stage1})--(\ref{eq:gating-stepgate}).

\paragraph{Causal masking.}
During training we use the HASS causal mask, adjusted so that each replaced position only attends to states available at its own timestep (no look-ahead).

\paragraph{Progressive replacement.}
Fig.~\ref{fig:attn} illustrates the unrolled rollout.
At draft depth $j{=}1$, the context uses target (teacher) features for all past positions (EAGLE-style).
At draft depth $j{\ge}2$, the most recent $(j{-}1)$ positions in the attention context are progressively \emph{replaced} by draft features produced at earlier depths, while earlier positions remain teacher features.
At each depth $j$, we inject the corresponding SPE $\mathbf{s}_j$ for the currently replaced window and queries, while providing per-token IPE to all positions.

\subsection{Inference}

At decoding time, the draft model augments inputs with the corresponding IPE and the SPE indexed by the \emph{current draft depth}.
It then proposes a candidate tree whose maximum depth is bounded by the fixed \emph{speculation depth} $B$ (following \cite{li2024eagle2}).
The target LLM verifies all candidates in one batch and commits the longest accepted prefix, advancing the output by multiple tokens per verification round.

PAD-Rec adds minimal overhead to the draft: two embedding lookups (IPE/SPE) and lightweight scalar gates per token.
The additional parameters introduced by PAD-Rec are negligible compared with the draft backbone (e.g., about $0.01\%$ when $B{=}6$).

\section{Experiments} 
We comprehensively evaluate \textbf{PAD-Rec} on four real-world datasets, comparing its inference acceleration against strong speculative-decoding baselines. Beyond the main comparison, we conduct ablations to disentangle the contribution of our \emph{item/step position embeddings} and the \emph{gating mechanisms}, and we analyze key hyperparameters that influence speedup.

To assess PAD-Rec for LLM-based generative list-wise recommendation, we answer the following research questions:

\begin{itemize}[leftmargin=*]
    \item \textbf{RQ1:} How does PAD-Rec compare with state-of-the-art SD methods in terms of wall-clock speedup, accepted length, and recommendation recall?
    \item \textbf{RQ2:} What are the individual and joint effects of item position embeddings, step position embeddings, and gating on PAD-Rec's inference acceleration?
    \item \textbf{RQ3:} How do hyper-parameters like speculation depth affect PAD-Rec’s efficiency?
    \item \textbf{RQ4:} How well does PAD-Rec scale across different backbone model sizes?
\end{itemize}

\subsection{Experimental Setup}

\subsubsection{Data Preparation.}

We evaluate PAD-Rec on four real-world datasets drawn from two widely used recommendation benchmarks.
The dataset statistics are summarized in Table~\ref{tab:statistics}.

\begin{itemize}[leftmargin=*]
    \item \textbf{Beauty}, \textbf{Instruments}, and \textbf{Games} are three subsets of the Amazon Reviews corpus~\cite{ni2019justifying}, covering user interactions with beauty and personal care products, musical instruments and related accessories, and video games, respectively.
    \item \textbf{Yelp} is derived from the Yelp Open Dataset\footnote{\url{https://www.yelp.com/dataset}}, which records user reviews and interactions with local businesses such as restaurants and services.
\end{itemize}

\textbf{Preprocessing and sequence construction.}
Following \cite{wang2024learnable}, we first obtain item \emph{semantic embeddings} using a pretrained encoder that integrates semantic and collaborative signals, and then discretize items into semantic IDs as described in Sec.~\ref{sec2.1}. User interactions are sorted chronologically, and users with fewer than 11 interactions are removed. For each remaining user, the most recent $10$ items are used as the \emph{target list}, while all earlier interactions constitute the \emph{history}.

\textbf{Train/validation/test splits.}
We split users into training, validation, and test sets with an 8:1:1 ratio at the user level. This ensures that all interactions of a given user are assigned to the same split, thereby preventing information leakage across data partitions.

\textbf{Speculative-decoding data.}
We first train the target LLM-based list-wise recommender following LC-Rec \cite{zheng2024adapting}. Its hidden features on the instruction-augmented token stream are then used to supervise the draft model.

\begin{table}[t]
\centering
\caption{Dataset statistics.}
\label{tab:statistics}
\setlength{\tabcolsep}{2pt}
\renewcommand{\arraystretch}{1.1}
\footnotesize
\begin{tabular}{l
                S[table-format=5]
                S[table-format=6]
                S[table-format=5]
                S[table-format=5]}
\toprule
\textbf{Dataset} & \textbf{\#Items} & \textbf{\#Inter.} & \textbf{\#Seqs} & \textbf{\#SD Seqs} \\
\midrule
Beauty      & 12101 & 198504 & 22363 & 6562 \\
Instruments &  9922 & 206153 & 24772 & 5753 \\
Games      & 17332 & 342329 & 49156 & 7721 \\
Yelp &  20033 & 316354 & 30431 & 13516 \\
\bottomrule
\end{tabular}

\vspace{2pt}
\footnotesize \emph{Note.} “\#Items”, “\#Inter.”, and “\#Seqs” are counted after preprocessing; “\#SD Seqs” denotes sequences used for SD training/evaluation.
\end{table}

\subsubsection{Baselines.}
We compare our proposed PAD-Rec against several strong SD baselines.

\begin{itemize}[leftmargin=*]
    \item \textbf{EAGLE-2.~\cite{li2024eagle2}} It uses a \emph{dynamic draft tree} with feature-level autoregression to raise verification acceptance and improve end-to-end speed.
    \item \textbf{HASS.~\cite{zhang2025learning}} It uses \emph{multi-aligning-step training} to match decoding-time context and improve acceptance.
    \item \textbf{FSPAD.~\cite{gui2024boosting}} It uses \emph{feature sampling} and \emph{partial alignment distillation} to reduce feature uncertainty and stabilize alignment, improving candidate quality and speed.
    \item \textbf{GRIFFIN.~\cite{hu2025griffin}} It uses \emph{token-alignable training} with \emph{token-guided fusion} to mitigate train–decode misalignment and improve acceptance length.
\end{itemize}

All baselines share the same target LLM and evaluation protocol; we adopt their official configurations for a fair comparison.

\begin{table*}[t]
\centering
\caption{Performance comparison between PAD-Rec and baselines on four datasets.}

\label{tab:main_speedup}
\setlength{\tabcolsep}{3pt}
\renewcommand{\arraystretch}{1.15}

\begin{tabular*}{\textwidth}{@{\extracolsep{\fill}} l p{3.2cm} c c c c c c c c}
\toprule
\textbf{Dataset} & \textbf{Model}
& \multicolumn{4}{c}{\textbf{temp=0}}
& \multicolumn{4}{c}{\textbf{temp=0.5}} \\
\cmidrule(lr){3-6}\cmidrule(lr){7-10}
& & \textbf{Speedup} & \textbf{$\tau$} & \textbf{Recall@10} & \textbf{NDCG@10}
  & \textbf{Speedup} & \textbf{$\tau$} & \textbf{Recall@10} & \textbf{NDCG@10} \\

\midrule

\multirow{6}{*}{\textbf{Beauty}} & \textbf{Target LLM}
& -- & -- & 0.0486 & 0.0593
& -- & -- & 0.0569 & 0.0629 \\
\cmidrule(lr){2-10}
& \textbf{EAGLE-2}
& 3.00 & 6.83 & 0.0480\down{0.0006} & 0.0588\down{0.0005}
& \underline{2.32} & 5.65 & 0.0531\down{0.0038} & 0.0582\down{0.0047} \\
& \textbf{HASS}
& \underline{3.04} & 6.87 & 0.0483\down{0.0003} & 0.0591\down{0.0002}
& 2.30 & 5.59 & 0.0563\down{0.0006} & 0.0628\down{0.0001} \\
& \textbf{FSPAD}
& 2.83 & 7.21 & 0.0485\down{0.0001} & 0.0592\down{0.0001}
& 2.31 & 5.96 & 0.0541\down{0.0028} & 0.0603\down{0.0026} \\
& \textbf{GRIFFIN}
& 2.82 & \textbf{7.83} & 0.0480\down{0.0006} & 0.0588\down{0.0005}
& 2.27 & \textbf{6.45} & 0.0523\down{0.0046} & 0.0593\down{0.0036} \\
& \textbf{PAD-Rec (ours)}
& \textbf{3.07} & \underline{7.35} & 0.0486\same{0.0000} & 0.0593\same{0.0000}
& \textbf{2.37} & \underline{6.22} & 0.0546\down{0.0023} & 0.0619\down{0.0010} \\

\midrule
\multirow{6}{*}{\textbf{Instruments}} & \textbf{Target LLM}
& -- & -- & 0.0337 & 0.0462
& -- & -- & 0.0323 & 0.0382 \\
\cmidrule(lr){2-10}
& \textbf{EAGLE-2}
& 2.39 & 4.55 & 0.0337\same{0.0000} & 0.0462\same{0.0000}
& 1.71 & 3.88 & 0.0336\up{0.0013} & 0.0411\up{0.0029} \\
& \textbf{HASS}
& 2.76 & 5.75 & 0.0339\up{0.0002} & 0.0463\up{0.0001}
& 2.24 & 5.01 & 0.0298\down{0.0025} & 0.0383\up{0.0001} \\
& \textbf{FSPAD}
& 3.01 & 7.13 & 0.0339\up{0.0002} & 0.0463\up{0.0001}
& 2.35 & 6.22 & 0.0322\down{0.0001} & 0.0390\up{0.0008} \\
& \textbf{GRIFFIN}
& \underline{3.03} & \textbf{7.89} & 0.0351\up{0.0014} & 0.0476\up{0.0014}
& \underline{2.39} & \textbf{6.81} & 0.0341\up{0.0018} & 0.0400\up{0.0018} \\
& \textbf{PAD-Rec (ours)}
& \textbf{3.15} & \underline{7.43} & 0.0337\same{0.0000} & 0.0462\same{0.0000}
& \textbf{2.44} & \underline{6.42} & 0.0291\down{0.0032} & 0.0376\down{0.0006} \\

\midrule
\multirow{6}{*}{\textbf{Games}} & \textbf{Target LLM}
& -- & -- & 0.0182 & 0.0221
& -- & -- & 0.0206 & 0.0220 \\
\cmidrule(lr){2-10}
& \textbf{EAGLE-2}
& 2.82 & 7.00 & 0.0182\same{0.0000} & 0.0220\down{0.0001}
& 2.17 & 5.66 & 0.0188\down{0.0018} & 0.0202\down{0.0018} \\
& \textbf{HASS}
& \underline{2.88} & 7.15 & 0.0184\down{0.0002} & 0.0222\up{0.0001}
& 2.21 & 5.93 & 0.0213\up{0.0007} & 0.0242\up{0.0022} \\
& \textbf{FSPAD}
& \textbf{2.94} & \underline{7.28} & 0.0182\same{0.0000} & 0.0220\down{0.0001}
& \underline{2.28} & \underline{6.13} & 0.0197\down{0.0009} & 0.0229\up{0.0009} \\
& \textbf{GRIFFIN}
& 2.72 & \textbf{8.00} & 0.0180\down{0.0002} & 0.0217\down{0.0004}
& 2.18 & \textbf{6.41} & 0.0201\down{0.0005} & 0.0217\down{0.0003} \\
& \textbf{PAD-Rec (ours)}
& \textbf{2.94} & \underline{7.28} & 0.0182\same{0.0000} & 0.0221\same{0.0000}
& \textbf{2.29} & 6.06 & 0.0201\down{0.0005} & 0.0221\up{0.0001} \\

\midrule
\multirow{6}{*}{\textbf{Yelp}} & \textbf{Target LLM}
& -- & -- & 0.0055 & 0.0075
& -- & -- & 0.0085 & 0.0088 \\
\cmidrule(lr){2-10}
& \textbf{EAGLE-2}
& 2.10 & 5.58 & 0.0053\down{0.0002} & 0.0074\down{0.0001}
& 1.68 & 4.80 & 0.0076\down{0.0009} & 0.0079\down{0.0009} \\
& \textbf{HASS}
& 2.29 & 5.84 & 0.0053\down{0.0002} & 0.0074\down{0.0001}
& 1.82 & 4.47 & 0.0076\down{0.0009} & 0.0073\down{0.0015} \\
& \textbf{FSPAD}
& 2.31 & \textbf{6.72} & 0.0053\down{0.0002} & 0.0074\down{0.0001}
& 2.06 & \underline{5.31} & 0.0083\down{0.0002} & 0.0079\down{0.0009} \\
& \textbf{GRIFFIN}
& \underline{2.34} & \underline{6.43} & 0.0052\down{0.0003} & 0.0070\down{0.0005}
& \underline{2.13} & \textbf{5.32} & 0.0082\down{0.0003} & 0.0086\down{0.0002} \\
& \textbf{PAD-Rec (ours)}
& \textbf{2.36} & 6.08 & 0.0054\down{0.0001} & 0.0074\down{0.0001}
& \textbf{2.20} & 4.77 & 0.0087\up{0.0002} & 0.0090\up{0.0002} \\

\bottomrule
\end{tabular*}
\vspace{2pt}
\begin{minipage}{0.98\linewidth}
\footnotesize
\textit{Note.} Results are reported on \textbf{Beauty}, \textbf{Instruments}, \textbf{Games}, and \textbf{Yelp}. For acceleration metrics, the best results are shown in \textbf{bold} and the second-best results are underlined. For \textit{Recall@10} and \textit{NDCG@10}, colored arrows indicate deviations from the target LLM: \textcolor{red}{$\uparrow$} denotes an increase and \textcolor{blue}{$\downarrow$} denotes a decrease, followed by the absolute change.
\end{minipage}

\end{table*}

\subsubsection{Evaluation Metrics.}

Following prior SD work, we report two efficiency metrics and two recommendation-quality metrics.
\begin{itemize}[leftmargin=*]
    \item \textbf{Wall-clock speedup} measures end-to-end acceleration, defined as the ratio between the inference time of standard autoregressive decoding and that of the speculative decoding pipeline under identical decoding settings.

    \item \textbf{Average accepted length ($\tau$)} measures the average number of drafted tokens accepted per verification step. A larger $\tau$ implies fewer target-model invocations and thus higher potential speedup.

    \item \textbf{Recall@10} measures recommendation accuracy, computed as $|\mathcal{G}\cap \hat{\mathcal{G}}|/|\mathcal{G}|$ for a ground-truth list $\mathcal{G}$ of size 10 and the generated list $\hat{\mathcal{G}}$, and averaged over all users.
    
    \item \textbf{NDCG@10} (Normalized Discounted Cumulative Gain) evaluates ranking quality by accounting for the positions of correctly recommended items. 
    We compute NDCG@10 between $\hat{\mathcal{G}}$ and $\mathcal{G}$, normalize by the ideal DCG, and report the average across users. 
    Higher NDCG@10 indicates better-ranked recommendations.

\end{itemize}

\subsubsection{Implementation Details.}
For the main comparison, we implement PAD-Rec on top of LC-Rec with Llama-3.2-1B-Instruct~\cite{dubey2024llama} as the target LLM; the model-scaling study further considers Llama-3.2-3B-Instruct and Llama-3-8B-Instruct. We adapt LC-Rec to our list-wise setting by formatting each supervision target as an ordered top-10 list of semantic-ID item tuples and training the model to autoregressively generate the flattened list token stream. We first fine-tune this list-wise LC-Rec model with full-parameter tuning to obtain the target model. The target LLM is then frozen, and only the lightweight draft module is trained. PAD-Rec keeps the target model and draft--then--verify framework unchanged, and only augments the draft-side representations with IPE/SPE and their gating parameters.

For a fair comparison, all SD baselines and PAD-Rec use the same target LLM, datasets, decoding length, evaluation metrics, and EAGLE-2-style tree-based verification framework. We sweep the learning rate in \{1e{-4},\,5e{-4},\,1e{-3}\}, set the speculation depth \(B\) to \(6\), and use a tree width of \(10\) at inference time. All reported results use the best validation setting under these choices. The draft module is built from a single Transformer layer of the same backbone; PAD-Rec keeps the target model and tree-based verification unchanged and only adds IPE/SPE tables and gating parameters on the draft side.

For the system protocol, speculative training uses bf16 mixed precision. Experiments with the 1B target model are trained on a single NVIDIA RTX 3090 GPU, while the 3B and 8B scaling experiments are trained on four NVIDIA A100 GPUs. All latency measurements are conducted on NVIDIA RTX 3090 GPUs in fp16 with batch size 1. We perform three warmup generations and exclude them from all reported numbers. During timing, CUDA is synchronized immediately before and after each generation call. The timed region includes draft proposal construction and target-model verification for SD methods, and the standard generation loop for the autoregressive target LLM; tokenization, post-decoding, file I/O, and metric computation are excluded. Both speculative and autoregressive decoding use the target model with KV cache enabled.

\subsection{Overall Performance (RQ1)}

We evaluate all methods on four datasets, namely \emph{Beauty}, \emph{Instruments}, \emph{Games}, and \emph{Yelp}, under identical inference configurations.
Table~\ref{tab:main_speedup} reports wall-clock speedup, average accepted length $\tau$, and recommendation quality measured by Recall@10 and NDCG@10 under two sampling temperatures. At temp$=\!0$, PAD-Rec matches or nearly matches the target LLM's recommendation metrics; at temp$=\!0.5$, it largely preserves recommendation quality with limited absolute fluctuations under stochastic sampling.
To contextualize the reported speedups in real time, Table~\ref{tab:naive_latency} reports the absolute autoregressive decoding latency of the target LLM in ms/query under the 1B setting. The latency is computed as \(\sum \texttt{wall\_time}/N\), where \(N\) is the number of evaluation queries and each query generates a top-10 recommendation list with approximately 59 output tokens. Since speedup in Table~\ref{tab:main_speedup} is measured relative to the same target-LLM baseline, the corresponding PAD-Rec latency can be estimated by dividing the target-LLM latency by the reported speedup.

\begin{table}[t]
\centering
\caption{Naive target-LLM decoding latency.}
\label{tab:naive_latency}
\setlength{\tabcolsep}{5pt}
\renewcommand{\arraystretch}{1.08}
\footnotesize
\begin{tabular}{l c c c}
\toprule
\textbf{Dataset} & \textbf{\#Queries} & \textbf{temp=0} & \textbf{temp=0.5} \\
\midrule
\textbf{Beauty} & 652 & 693.55 & 725.64 \\
\textbf{Instruments} & 588 & 676.70 & 725.43 \\
\textbf{Games} & 757 & 1054.68 & 1060.07 \\
\textbf{Yelp} & 1311 & 882.77 & 917.73 \\
\bottomrule
\end{tabular}

\vspace{2pt}
\begin{minipage}{0.95\linewidth}
\footnotesize
\textit{Note.} Latency is reported as mean wall-clock decoding time per query (ms/query) for the 1B target LLM with batch size 1.
\end{minipage}
\end{table}

\begin{itemize}[leftmargin=*]
\item \textbf{HASS vs.\ EAGLE-2.}
Across all datasets, HASS consistently outperforms EAGLE-2 in terms of accepted length $\tau$ and wall-clock speedup.
This trend is observed on \emph{Beauty}, \emph{Instruments}, and \emph{Games}, and also holds on the smaller-scale \emph{Yelp} dataset.
The results indicate that exposing the draft model to multi-step speculative contexts during training better aligns it with the decoding-time distribution, resulting in fewer verification rejections and higher end-to-end efficiency, while maintaining comparable Recall@10 and NDCG@10.

\item \textbf{FSPAD and GRIFFIN.}
Both FSPAD and GRIFFIN improve draft--target alignment and yield longer accepted prefixes, with GRIFFIN typically achieving the largest $\tau$.
However, GRIFFIN’s token-guided fusion introduces additional MLP layers during draft-time inference, increasing memory traffic and latency; as a result, gains in $\tau$ do not consistently translate into higher wall-clock speedup.
Across datasets, Recall@10 and NDCG@10 remain comparable to the Target LLM, indicating that these methods mainly affect efficiency rather than recommendation quality.

\item \textbf{PAD-Rec (ours).}
PAD-Rec achieves the highest or near-highest wall-clock speedup across all datasets and temperature settings.
At temp$=\!0$, PAD-Rec reaches \textbf{3.07$\times$} speedup on \emph{Beauty} and \textbf{3.15$\times$} on \emph{Instruments}, with similarly strong acceleration on \emph{Games} and \emph{Yelp}.
These gains are obtained without introducing heavy fusion modules or changing the target-model verification path.
Instead, PAD-Rec relies on two lightweight positional signals.
IPE provides within-item slot awareness for semantic-ID tuples, while SPE conditions the draft on the current speculative depth, allowing the model to adapt its behavior as decoding progresses.
Both signals are integrated via simple gating mechanisms, achieving better end-to-end speedup with competitive accepted length and lower draft-side overhead.

\item \textbf{Effect of temperature.}
Sampling temperature has a systematic impact on speculative decoding.
At temp$=\!0$, decoding is deterministic and closely follows the target model’s top-1 trajectory, resulting in strong proposal--target alignment; consequently, all SD variants achieve long accepted prefixes, stable Recall@10/NDCG@10, and near-maximal wall-clock speedup.
When the temperature increases to temp$=\!0.5$, speedup consistently decreases and variability in Recall@10 and NDCG@10 becomes more pronounced across methods.
This behavior is expected, as stochastic sampling increases the divergence between draft proposals and target continuations and shifts generation away from the regime used during draft training, leading to earlier verification termination and shorter accepted lengths $\tau$.
Moreover, under higher stochasticity, the space of plausible continuations expands rapidly, while the finite draft tree can only cover a limited subset, further increasing the likelihood of early rejection during verification.
Overall, these results highlight an inherent efficiency--robustness trade-off in speculative decoding; within this trade-off, PAD-Rec consistently maintains strong acceleration while largely preserving recommendation quality under increased sampling stochasticity.

\end{itemize}

\begin{figure}[t]
\setlength{\abovecaptionskip}{0pt}
\setlength{\belowcaptionskip}{-2pt}
\centering
\includegraphics[width=\linewidth]{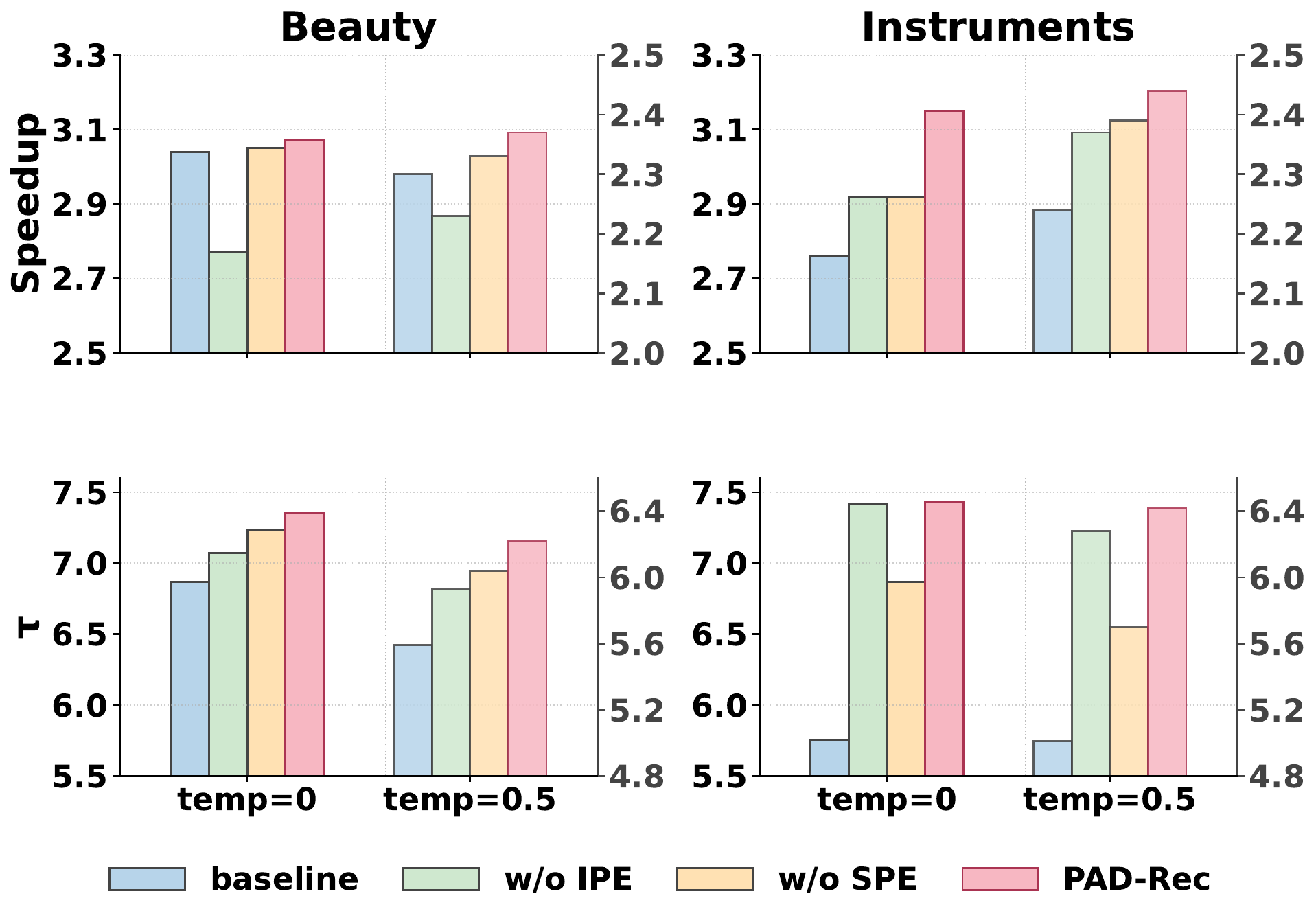}
\caption{Embedding ablation on \emph{speedup} and \emph{accepted length} $\tau$ on {Beauty} and {Instruments}. In each subplot, temp$=\!0$ is read from the left $y$-axis, while temp$=\!0.5$ is read from the right $y$-axis.}
\label{fig:ablation1}
\end{figure}

\subsection{Ablation Study (RQ2)}

\subsubsection{Embedding ablation.}

To evaluate the effects of IPE and SPE, we compare \textbf{PAD-Rec} under the same SD setting with its variants and report wall-clock \emph{speedup} and accepted length $\tau$.
Due to space constraints, we present results on two representative datasets, {Beauty} and {Instruments}, and conduct the remaining analytical experiments on the same datasets, with similar trends observed on the others.
The ablations comprise \emph{Baseline (no PAD-Rec)}, \emph{PAD-Rec w/o IPE}, \emph{PAD-Rec w/o SPE}, and \emph{PAD-Rec (full)}.
In Fig.~\ref{fig:ablation1}, the two temperatures use separate $y$-axes for readability: temp$=\!0$ corresponds to the left axis and temp$=\!0.5$ corresponds to the right axis.
From Fig.~\ref{fig:ablation1} we observe:

\begin{itemize}[leftmargin=*]
\item \textit{Overall acceleration.} The full PAD-Rec yields the strongest or near-strongest speedup across datasets and temperatures while keeping $\tau$ among the top two, indicating more tokens verified per target call without notable complexity.
\item \textit{Effect of IPE.} Removing IPE reduces speedup on structured outputs (semantic-ID tuples), showing that explicit slot cues help the draft model form proposals that verify more readily.
\item \textit{Effect of SPE.} Removing SPE is most harmful at temp$=\!0.5$, where deeper draft steps are noisier; step-aware conditioning improves stability at depth, preserving speedup.
\item \textit{Complementarity.} IPE (item-structure) and SPE (depth-adaptive) are complementary: the full model consistently outperforms either single ablation.
\end{itemize}

\noindent\textbf{Takeaway.} Lightweight slot and step signals—modulated by simple gates—improve draft–target alignment and deliver consistent acceleration with minimal overhead.

\subsubsection{Gating ablation.}

\begin{figure}[!t]
\setlength{\abovecaptionskip}{0pt}
\setlength{\belowcaptionskip}{0pt}
\centering
\includegraphics[width=\linewidth]{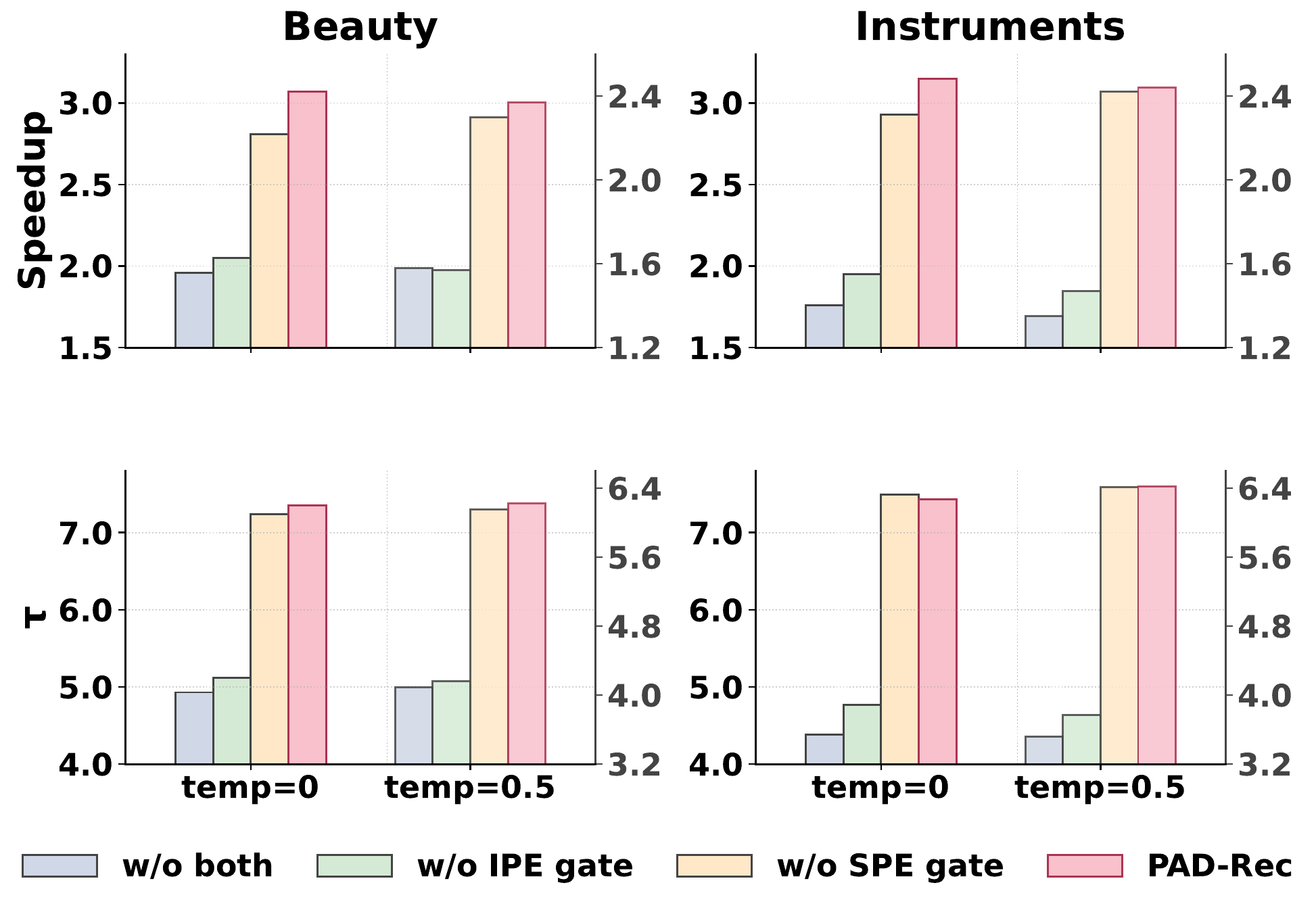}
\caption{Gate ablation on \emph{speedup} and accepted length $\tau$ on \textbf{Beauty} and \textbf{Instruments}. In each subplot, temp$=\!0$ is read from the left $y$-axis, while temp$=\!0.5$ is read from the right $y$-axis.}
\label{fig:ablation_gate}
\end{figure}

To assess the necessity of the two gates modulating positional signals, we conduct another ablation study. We compare four variants: \emph{w/o both gates} (disabling both), \emph{w/o IPE gate} (disabling only the item gate), \emph{w/o SPE gate} (disabling only the step gate), and the full model \emph{PAD-Rec}. As in Fig.~\ref{fig:ablation1}, Fig.~\ref{fig:ablation_gate} uses the left $y$-axis for temp$=\!0$ and the right $y$-axis for temp$=\!0.5$. The results show:

\begin{itemize}[leftmargin=*]
\item \textit{Removing both gates hurts most.} Without any gating, speedup drops sharply across all settings (e.g., $\sim\!1.96{\times}$ on \textit{Beauty} and $\sim\!1.76{\times}$ on \textit{Instruments} at temp$=\!0$), accompanied by the lowest $\tau$. This shows that ungated addition of position signals destabilizes proposals and lowers acceptance.
\item \textit{Item gate is crucial.} Disabling only the {IPE gate} (\textit{w/o IPE gate}) remains far from the full model, close to the \textit{w/o both} variant. This indicates the item gate is essential to scale slot cues properly and avoid overwhelming token semantics.
\item \textit{Step gate provides consistent gains.} Disabling only the {SPE gate} (\textit{w/o SPE gate}) retains most of the performance (e.g., speedup $2.81{\times}$ vs.\ $3.07{\times}$ on \textit{Beauty}; $2.93{\times}$ vs.\ $3.15{\times}$ on \textit{Instruments} at temp$=\!0$), with $\tau$ very close to the full model. Thus, the step gate yields \emph{modest but consistent} improvements by fine-tuning depth-wise conditioning rather than being the primary driver.
\item \textit{Full model is best.} {PAD-Rec (full)} attains the highest speedup in all settings and the top (or near-top) $\tau$, confirming that \emph{both} gates together deliver the most reliable acceleration.
\end{itemize}

\noindent\textbf{Takeaway.} The item gate serves as the primary stabilizer for slot-aware cues, while the step gate provides a lighter depth-adaptive adjustment; together they maximize acceptance and end-to-end acceleration with negligible overhead.

\subsection{Hyper-parameter Analysis (RQ3)}

\begin{figure}[t]
\setlength{\abovecaptionskip}{0pt}
\setlength{\belowcaptionskip}{-5pt}
\centering
\includegraphics[width=\linewidth]{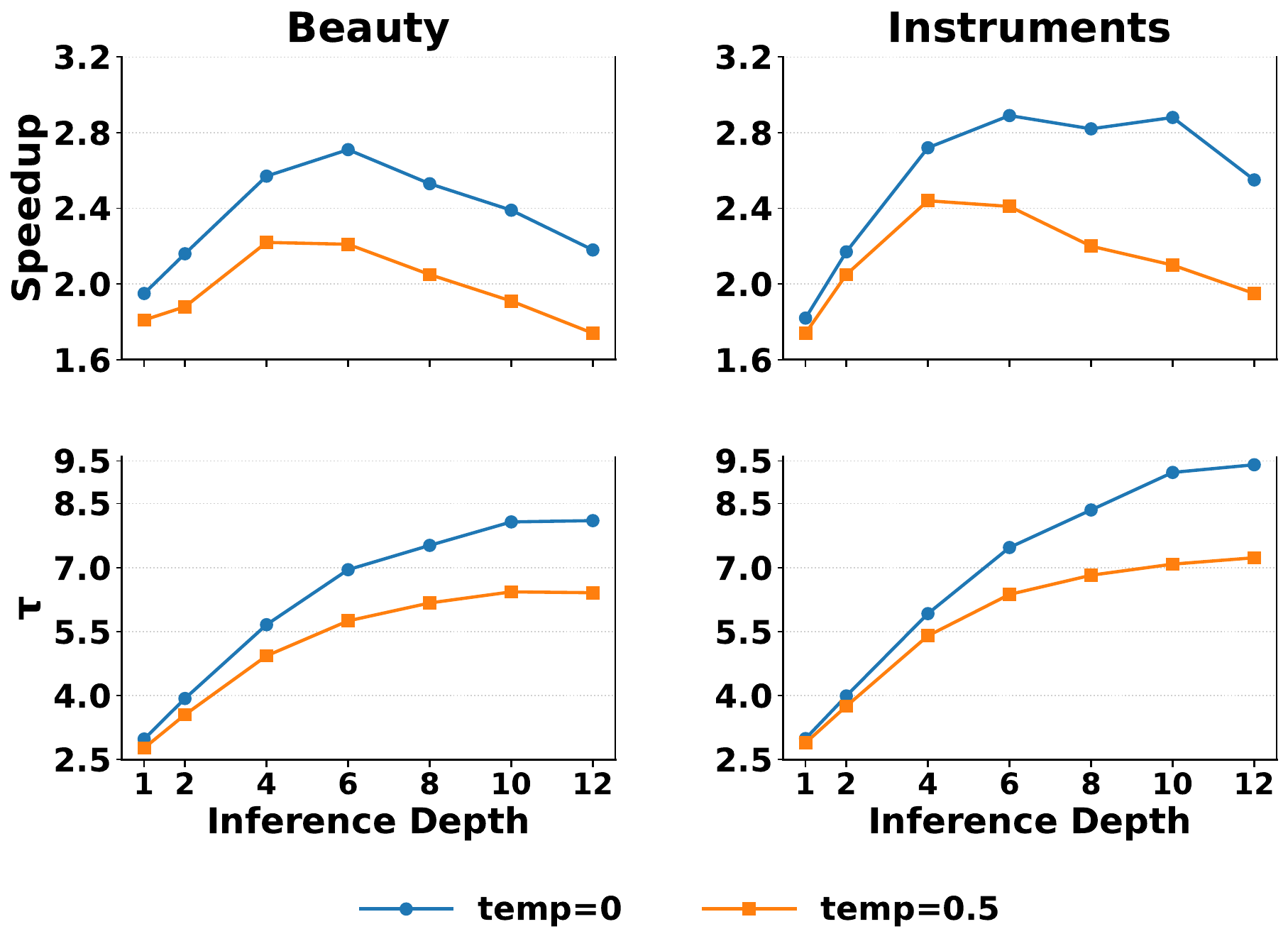}
\caption{Effect of speculation depth $B_{\text{test}}$ (with $B_{\text{train}}{=}12$) on \emph{speedup} and accepted length $\tau$ at temp$=\!0$ and $0.5$ on \textbf{Beauty} and \textbf{Instruments}.}
\label{fig:hyper_depth}
\end{figure}
\noindent\textbf{Speculation depth.}
We study the effect of inference speculation depth $B_{\text{test}}$ while fixing the training speculation depth at $B_{\text{train}}{=}12$ (to fully cover step positions for SPE). We vary $B_{\text{test}} \in \{1, 2, 4, 6, 8, 10, 12\}$ across both datasets and report \emph{speedup} and \emph{accepted length} $\tau$ (Fig.~\ref{fig:hyper_depth}). The main observations are as follows:

\begin{itemize}[leftmargin=*]
\item \textit{Growth of $\tau$ with depth.} For both datasets and temperatures, $\tau$ increases monotonically as $B_{\text{test}}$ grows. Initially, it increases linearly up to $B=6$, while speedup also rises significantly with $\tau$. However, after $B=6$, $\tau$ grows more slowly, converging to around 7, and speedup starts to drop.

\item \textit{Speedup peaks at moderate depths.} Speedup follows a unimodal trend: it increases with depth, peaks at moderate values (typically around $B_{\text{test}} \approx 4$–$6$), and then decreases as the depth continues to grow. This reflects the trade-off between (i) longer accepted prefixes (larger $\tau$) and (ii) the overhead and diminishing returns in acceptance when proposing too deep a draft. In our structured outputs, each item consists of a small, fixed number of semantic-ID tokens plus separators, and moderate $B_{\text{test}}$ already spans roughly one item, which is generally the sweet spot for end-to-end latency.

\item \textit{Across temperatures.} The trends above hold for both temp$=\!0$ and temp$=\!0.5$; quantitatively, temp$=\!0$ is slightly higher in speedup at its peak, yielding more stable, longer accepted prefixes at the same $B_{\text{test}}$.

\item \textit{Practical choice of depth.} Although we trained with $B_{\text{train}}{=}12$ to cover all step indices for SPE, the optimal \emph{inference} depth is smaller (e.g., $B_{\text{test}}\!\approx\!4$–$6$). Therefore, for deployment, we default to a moderate speculation depth of $B_{\text{test}}{=}6$ to balance acceptance length and verification/branching overhead.
\end{itemize}

\subsection{Scaling Analysis (RQ4)}

\begin{figure}[t]
\setlength{\abovecaptionskip}{0pt}
\setlength{\belowcaptionskip}{-4pt}
    \centering
    \includegraphics[width=\columnwidth]{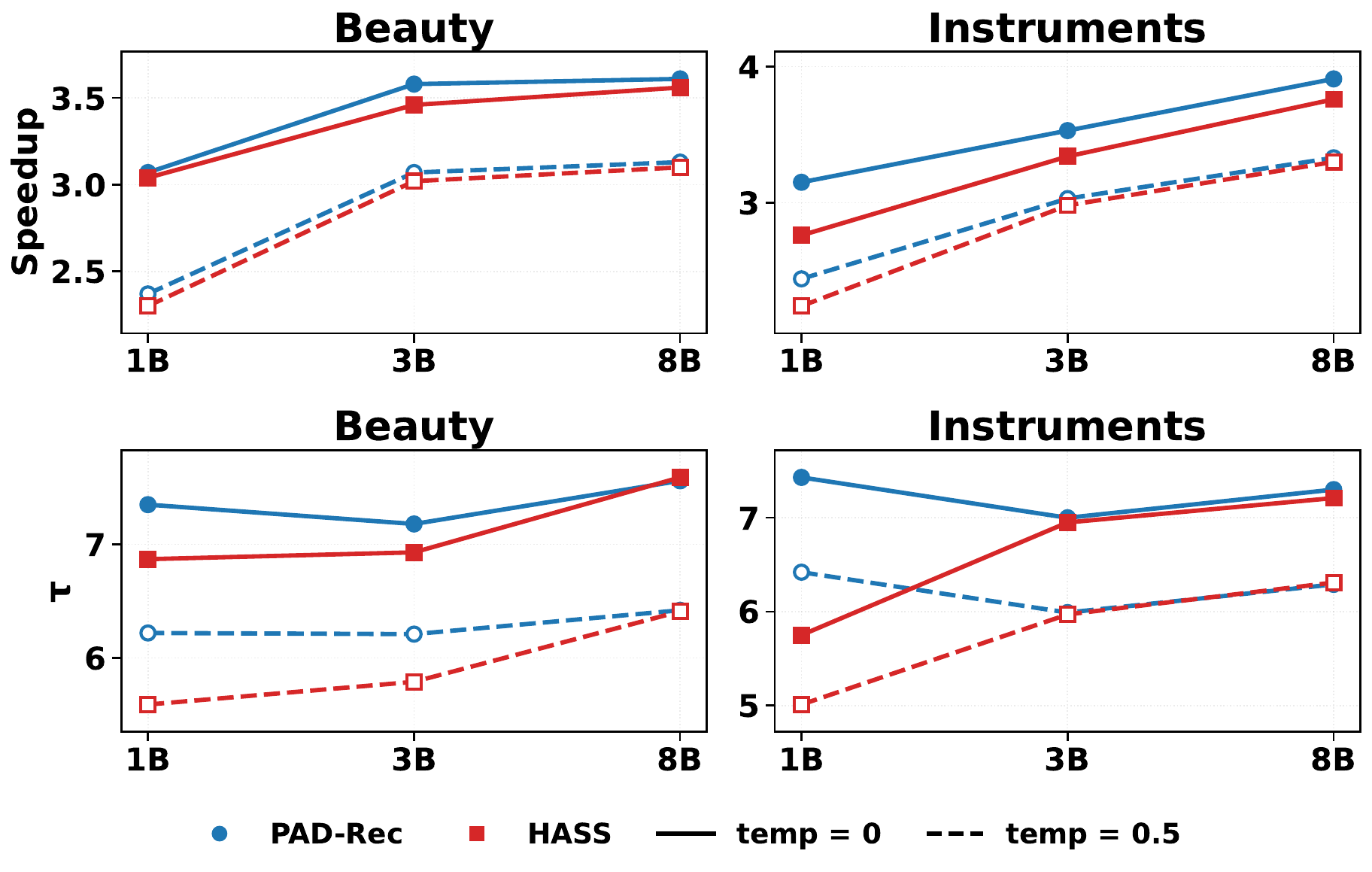}
    \caption{\textbf{Scaling analysis} on \emph{speedup} and accepted length $\mathbf{\tau}$ at temp$=\!0$ and $0.5$ on \textbf{Beauty} and \textbf{Instruments}.}
    \label{fig:size}
\end{figure}

\noindent\textbf{Model size.}
We analyze the scaling behavior of speculative decoding by varying the backbone model size and comparing PAD-Rec with HASS, while fixing the other hyper-parameters (with $B{=}6$).
Specifically, we evaluate three instruction-tuned backbones, \emph{Llama-3.2-1B-Instruct}, \emph{Llama-3.2-3B-Instruct}, and \emph{Llama-3-8B-Instruct}~\cite{dubey2024llama}.
For each backbone, we first perform recommendation fine-tuning under the same protocol, and then report inference \emph{speedup} and accepted length $\tau$ at temp$=\!0$ and $0.5$ on both {Beauty} and {Instruments} (see Fig.~\ref{fig:size}).

The main observations are summarized as follows:

\begin{itemize}[leftmargin=*]
\item \textit{Speedup increases with model size.}
Across both datasets and temperatures, speedup consistently improves as the backbone scales from 1B to 8B.
This indicates that speculative decoding benefits more from larger backbones, where the relative overhead of draft generation and verification becomes smaller.

\item \textit{Diminishing marginal gains.}
Although speedup grows monotonically with model size, the improvement from 3B to 8B is smaller than that from 1B to 3B, exhibiting a clear diminishing-return effect.
This suggests that scaling the backbone yields progressively smaller marginal benefits for speculative acceleration.

\item \textit{PAD-Rec vs. HASS.}
PAD-Rec consistently achieves higher speedup than HASS across all model sizes and temperatures.
This demonstrates that PAD-Rec can more effectively exploit speculative acceptance for acceleration, and its advantage remains stable as the model scales.

\item \textit{Accepted length $\tau$.}
The accepted length $\tau$ increases mildly with model size and remains relatively stable compared to speedup.
This implies that scaling behavior is mainly driven by efficiency gains rather than substantial changes in acceptance behavior.
\end{itemize}
\section{Conclusion}

In this work, we study speculative decoding for LLM-based generative list-wise recommendation and highlight that effective acceleration must respect both the structured nature of item representations and the multi-step decoding process.
To this end, we propose PAD-Rec, a position-aware drafting module that augments the draft model with item position embeddings and step position embeddings, modulated by lightweight gates to better align draft proposals with structured recommendation outputs.
Extensive experiments on four real-world datasets demonstrate that PAD-Rec consistently achieves substantial inference acceleration—up to \textbf{3.1$\times$} wall-clock speedup—while incurring negligible overhead and largely preserving recommendation quality.
Further analyses show that PAD-Rec is robust across sampling temperatures, hyper-parameter choices, and backbone model scales, and consistently outperforms prior speculative decoding baselines.
These results indicate that incorporating structure-aware inductive biases into speculative decoding is a practical and effective approach for accelerating LLM-based generative recommendation systems.

\bibliographystyle{IEEEtran}
\bibliography{main}

\end{document}